\documentclass[a4,12pt]{article}

\usepackage{a4}

\usepackage{amsmath}
\usepackage{amssymb}
\usepackage{amsfonts}
\usepackage{bbold}
\usepackage{mathrsfs}
\usepackage{bm}
\usepackage[]{hyperref}
\usepackage{graphicx}
\usepackage{braket}
\usepackage{multirow}
\usepackage{cite}
\hypersetup{colorlinks=true,linkcolor=blue, citecolor=blue,urlcolor=blue}

\def\Pexp{\mathop{\rm Pexp}\nolimits}

\def\N{\mathop{\mathcal{N}}\nolimits}
\def\AdS{\mathop{\mathrm{AdS_5\times S^5}}\nolimits}

\def\tr{\mathop{\rm tr}\nolimits}

\allowdisplaybreaks

\newcommand{\ZZ}{\mathbb{Z}}

\def\N{\mathop{\mathcal{N}}\nolimits}
\newcommand{\nn}{\nonumber\\}

\def\Pexp{\mathop{\rm Pexp}\nolimits}

\begin{document}
\begin{titlepage}
\title{
\vspace{-1.5cm}
\begin{flushright}
{\normalsize Feb 2023}
\end{flushright}
\vspace{1.5cm}
\LARGE{Schur-like index of the Klebanov-Witten theory via the AdS/CFT correspondence}}
\author{Shota {\scshape Fujiwara\footnote{E-mail: shota.fujiwara@wits.ac.za}}
\\
\\
{\itshape Mandelstam Institute for Theoretical Physics, }
\\ {\itshape University of Witwatersrand, Johannesburg 2050, South Africa}}

\date{}
\maketitle
\thispagestyle{empty}

\begin{abstract}
We calculate an unrefined limit of the superconformal index (the Schur-like index) of the Klebanov-Witten theory 
from the dual AdS side by using the recently developed giant graviton expansion method.
Our formula includes multiple sums of the contribution of the D3-branes wrapped on 
three-cycles in the internal space.
We numerically confirm the validity of our formula by comparing indices obtained by the formula
 with the gauge theory results.
\end{abstract}
\end{titlepage}
\section{Introduction}
Ever since the discovery of the AdS/CFT correspondence \cite{hep-th/9711200},
the duality has been providing us with new ways to analyze the superconformal field theories.
Among them the superconformal index is a very efficient tool
since the superconformal index is independent of the coupling.
Actually, a lot of work has been done for the superconformal index
in relation to the AdS/CFT correspondence.
For example, at the large $N$ limit the superconformal index of the
four-dimensional $\N=4$ $U(N)$ SYM theories 
were reproduced from type IIB supergravity on $\AdS$ \cite{hep-th/0510251}.

Recently, great progress has been achieved toward
the calculation of the index at the finite $N$ region
by using the method called the giant graviton expansion
\cite{arXiv:1904.09776,arXiv:1907.05660,arXiv:1911.10794,arXiv:2001.11667,arXiv:2007.05213,arXiv:2108.12090,arXiv:2110.14897,arXiv:2109.02545,arXiv:2204.09286,arXiv:2202.06897,arXiv:2205.14615,arXiv:2212.05408}.
See also \cite{hep-th/0003075,hep-th/0008016,hep-th/0010206,hep-th/0606087}
for references about the giant graviton itself.
For the AdS$_5\times S^5$ case, even at the finite $N$ 
the index was reproduced by including the contribution of D3-branes wrapping on
three-cycles in $S^5$ \cite{arXiv:1904.09776,arXiv:2108.12090}.
More precisely, the authors of \cite{arXiv:1904.09776}
study the single-wrapping D3-brane contribution to the index,
and the multiple-wrapping D3-branes contribution was also considered in \cite{arXiv:2108.12090}.
The author numerically checked the following giant graviton expansion formula for small $N$:
\begin{align}
\label{s5formula}
\frac{\mathcal{I}^{\mathrm{SYM}}_{N}}{\mathcal{I}_{\mathrm{KK}}^{S^5}}
=\sum_{n_1,n_2,n_3=0}^{\infty}\mathcal{I}_{(n_1,n_2,n_3)}.
\end{align}
In this equation $\mathcal{I}^{\mathrm{SYM}}_{N}$ denotes the superconformal index of the 
$\N=4$ $U(N)$ SYM theory and $\mathcal{I}_{\mathrm{KK}}^{S^5}$ is the Kaluza-Klein index of 
the dual AdS$_5\times S^5$ supergravity.
Also, $n_i$ ($i=1,2,3$) denotes the wrapping number of D3-branes 
for each three-cycle and $\mathcal{I}_{(n_1,n_2,n_3)}$ is 
a superconformal index of the D3-brane system with the wrapping numbers $(n_1,n_2,n_3)$.
Note that this formula contains multiple sums of the D3-brane configurations,
but the authors of \cite{arXiv:2109.02545} show a similar formula with only a single sum.
The relation of these two formulae was studied in \cite{arXiv:2205.14615}.

From the success of the AdS$_5\times S^5$ case, it is natural to expect that similar relations hold
for other examples of the AdS/CFT correspondence. 
When $S^5$ is replaced by a five-dimensional Sasaki-Einstein manifold ($\mathrm{SE}_5$),
the corresponding gauge theory becomes a toric quiver gauge theory 
which is realized on D3-branes placed on the Calabi-Yau three-folds.
The superconformal indices for these toric cases
at the large $N$ limit were studied in
\cite{arXiv:1207.0573,arXiv:1304.6733}.
In addition, finite $N$ corrections
from the single-wrapping D3-brane were studied in \cite{arXiv:1911.10794}.
However, the multiple-wrapping D3-branes contribution
was not considered in that paper, and it is a remaining problem.
If we include the multiple-wrapping D3-branes
the natural extension of the formula (\ref{s5formula}) for the toric cases becomes
\begin{align}
\label{toricformula}
\frac{\mathcal{I}^{\mathrm{toric}}_{N}}{\mathcal{I}_{\mathrm{KK}}^{\mathrm{SE}_5}}
=\sum_{n_1,n_2,...,n_d=0}^{\infty}\mathcal{I}_{(n_1,n_2,...,n_d)},
\end{align}
where $d$ is the number of vertices of the toric diagram. 

In this paper, as a simple example of a $\mathrm{SE}_5$,
we especially study the $\mathrm{SE}_5=T^{1,1}$ case.
The dual 4d $\N=1$ $SU(N)_1\times SU(N)_2$ superconformal field theory
is known as the Klebanov-Witten theory \cite{hep-th/9807080}.
However, as discussed in \cite{arXiv:2108.12090}
there are some difficulties regarding how we determine the contours of
the gauge integral in $\mathcal{I}_{(n_1,n_2,...,n_d)}$.
To avoid this problem we instead consider an unrefined limit
of the superconformal index like the Schur limit of four-dimensional $\N=2$ theories
\cite{arXiv:1110.3740}.
(See Subsection \ref{Schursec} for our definition of the limit.)
If we take such a limit sometimes a nice simplification occurs and
the analysis becomes easier.
For example, giant graviton expansions with such Schur (like) limits were studied in the literature for
the $N=4$ $U(N)$SYM theory in \cite{arXiv:2001.11667}
and for the 6d $\N=(2,0)$ theory in\cite{arXiv:2007.05213}.
Then the purpose of this paper is to 
confirm the formula (\ref{toricformula}) for the unrefined index of the 
Klebanov-Witten theory.

\section{Index for the Klebanov-Witten theory}
Let us explain the definition of the superconformal index for the Klebanov-Witten theory.
Our convention follows from \cite{arXiv:1911.10794} and we will summarize shortly here.
For a detailed explanation of our convention see Section 2 and Subsection 4.1 of \cite{arXiv:1911.10794}.

\begin{figure}[htb]
\centering
\includegraphics[scale=0.5]{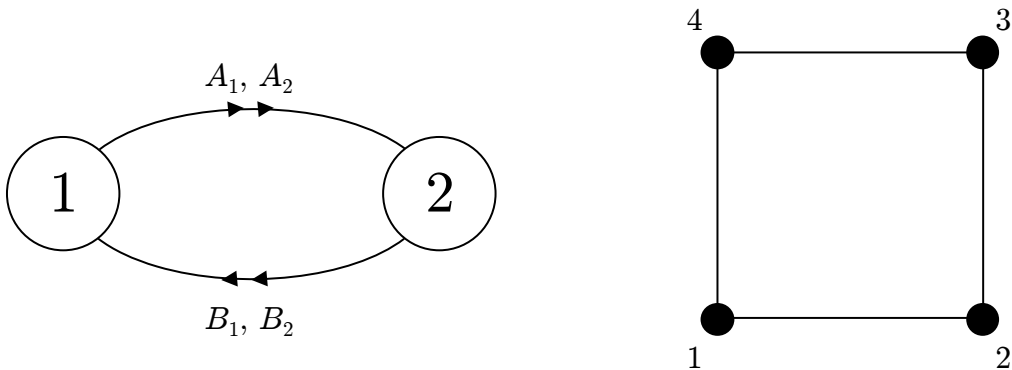}
\caption{The quiver diagram of the Klebanov-Witten theory and the toric diagram of $T^{1,1}$ are shown.}
\label{conitoric}
\end{figure}

The Klebanov-Witten theory is a 4d $\N=1$ $SU(N)_1\times SU(N)_2$ gauge theory described
by the quiver diagram shown in Figure \ref{conitoric}.
The dual gravity theory is type IIB superstring theory on AdS$_5\times T^{1,1}$.
The toric diagram of $T^{1,1}$ is also shown in Figure \ref{conitoric}.
We denote a supersymmetric three-cycle corresponding
to a vertex of the toric diagram as $S_I$ ($I=1\sim4$).

There are four independent charges $R_I$ ($I=1\sim4$) corresponding to the four vertices of the toric diagram.
We can determine the charge assignments of the fields by using the perfect matching
\cite{hep-th/0504110,hep-th/0506232,hep-th/0512240}.
Our charge assignments are summarized in Table \ref{chargetable}.
\begin{table}[htb]
\caption{The charge assignments for the Klebanov-Witten theory is shown.}
\vspace{-0.5cm}
\begin{align*}
\begin{array}{c|l|l|l|l|l|l|l|l}\hline
  &R_1&R_2&R_3&R_4&r_*&F_A&F_B&NB\\\hline
  A_1&  1& 0 &0  &  0&1/2&1&0&1\\
  A_2 & 0 & 1&0 & 0&1/2&0&1&-1\\	
  B_1 & 0 &0 &1 &0&1/2&-1&0&1 \\	
  B_2 &0  &0 &0 & 1&1/2&0&-1&-1\\\hline
\end{array}
\end{align*}
\label{chargetable}
\end{table}
Note that on the gravity side $R_I$ corresponds to the angular momentum
and a D3-brane wrapped on $S_I$ has $R_I=N$ and otherwise 0.
Then we define the superconformal index of the Klebanov-Witten theory by
\begin{align}
  \mathcal{I}^{\mathrm{KW}}=\tr\left[(-1)^{F}
  q^{3\overline{J}}y^{2J}v_1^{R_1}v_2^{R_2}v_3^{R_3}v_4^{R_4}\right], \quad  v_1v_2v_3v_4=q^3.
\label{defconiindexv}
\end{align}

It is also useful to use another basis for the charges.
As global symmetries, the Klebanov-Witten theory has an R-symmetry,
two $SU(2)$ mesonic symmetries, and also one baryonic symmetry.
We denote the superconformal $R$-charge obtained by the a-maximization \cite{hep-th/0304128} as 
$r_*$ and the Caltan generator of two $SU(2)$ as $F_A$ and $F_B$,
and the baryonic charge as $B$.
These generators can be expressed as a linear combination of $R_I$
and we take as follows:
\begin{align}
r_*=\frac{1}{2}(R_1+R_2+R_3+R_4),
\end{align}
\begin{align}
F_A=R_1-R_3,\qquad F_B=R_2-R_4,
\end{align}
\begin{align}
B=\frac{1}{N}(R_1-R_2+R_3-R_4).
\end{align}
We introduce new fugacities $u,v, \zeta$ corresponding to these generators.
In terms of these fugacities the index is written by 
\begin{align}
\label{defconiindex}
\mathcal{I}^{\mathrm{KW}}=\tr\left[(-1)^{F}
q^{\frac{3}{2} r_*+3 \overline{J}} y^{2J}u^{F_A}v^{F_B}\zeta^{B}\right],
\end{align}
where fugacities in this equation and the previous ones in (\ref{defconiindexv}) are related by
\begin{align}
v_1=q^{\frac{3}{4}} u \zeta^{\frac{1}{N}},\qquad
v_2=\frac{q^{\frac{3}{4}} v}{\zeta^{\frac{1}{N}}},\qquad
v_3=\frac{q^{\frac{3}{4}} \zeta^{\frac{1}{N}}}{u}, \qquad
v_4=\frac{q^{\frac{3}{4}}}{v\zeta^{\frac{1}{N}}}.
\end{align}
We will use this expression of the index in numerical calculation.

Notice that the baryonic charge is related to the wrapping number of D3-branes on the gravity side
\cite{hep-th/9808075}.
Let $n_I$ ($I=1\sim 4$) be the wrapping number of D3-branes for each cycle $S_I$.
We can easily find a D3-brane configuration 
with the wrapping numbers $(n_1,n_2,n_3,n_4)$ has
\begin{align}
B=n_1-n_2+n_3-n_4.
\end{align}

\section{Giant graviton expansion for the Klebanov-Witten theory}
Let us discuss the giant graviton expansion formula for the Klebanov-Witten theory.
The formula (\ref{toricformula}) now becomes
  \begin{align}
  \label{kwformula}
    \frac{\mathcal{I}^{\mathrm{KW}}_{N}}{\mathcal{I}_{\mathrm{KK}}^{T^{1,1}}}
    =\sum_{n_1,n_2,n_3,n_4=0}^{\infty}\mathcal{I}_{(n_1,n_2,n_3,n_4)},
  \end{align}
where $\mathcal{I}_\mathrm{KK}^{T^{1,1}}$ is the Kaluza-Klein contribution of 
supergravity on AdS$_5\times T^{1,1}$ \cite{hep-th/0602284}
and it is given by
\begin{align}
\mathcal{I}_{\mathrm{KK}}^{T^{1,1}}=\Pexp [i_{\mathrm{KK}}^{T^{1,1}}].
\end{align}
$\Pexp$ is the plethystic exponential given by
\begin{align}
\Pexp[f(x_i)]=\mathrm{exp}\left[\sum_{n=1}^{\infty}\frac{1}{n}f(x_i^n)\right],
\end{align}
and $i_{\mathrm{KK}}^{T^{1,1}}$ is the single-particle index of the Kaluza-Klein modes on $T^{1,1}$.
It is given by the following sum \cite{arXiv:1207.0573,arXiv:1304.6733}:
\begin{align}
i_{\mathrm{KK}}^{T^{1,1}}=\sum_{I=1}^{4}\frac{w_{I+\frac{1}{2}}}{1-w_{I+\frac{1}{2}}},
\end{align}
where $w_{I+\frac{1}{2}}$ $(I=1\sim 4)$ are fugacities associated with the edge 
between the vertices $I$ and $I+1$ of the toric diagram
\footnote{Here we treat $I$ as a cyclic variable $I\sim I+4$; for example,
$w_{4+\frac{1}{2}}=w_{\frac{1}{2}}$.}
and explicitly given by \cite{arXiv:1911.10794}
\begin{align}
w_{\frac{1}{2}}=v_2 v_3,\qquad w_{1+\frac{1}{2}}=v_3 v_4,
\qquad w_{2+\frac{1}{2}}=v_4 v_1,\qquad w_{3+\frac{1}{2}}=v_1 v_2.
\end{align}
On the right hand side of the the formula, $\mathcal{I}_{(n_1,n_2,n_3,n_4)}$ is the contribution of
the D3-branes with wrapping numbers $(n_1,n_2,n_3,n_4)$:
\begin{align}
\label{d3integral}
\mathcal{I}_{(n_1,n_2,n_3,n_4)}=v_1^{n_1}v_2^{n_2}v_3^{n_3}v_4^{n_4}
\int d\mu_1d\mu_2d\mu_3d\mu_4 \Pexp[i_{\mathrm{all}}].
\end{align}
\label{d3coni}
$d\mu_I$ is the Haar measure of $U(n_I)$ gauge group.
$i_{\mathrm{all}}$ is the single-particle index of the 
quiver theory on the D3-brane system given by 
\begin{align}
i_{\mathrm{all}}=\sum_{I=1}^{4}\left(
i^{v}_I \chi_{U(n_I)}^{\mathrm{adj}} 
+ i^{h}_{I,I+1} 
(\chi_{U(n_I)}^{\mathrm{fund}}\chi_{U(n_{I+1})}^{\overline{\mathrm{fund}}}
+\chi_{U(n_I)}^{\overline{\mathrm{fund}}}\chi_{U(n_{I+1})}^{\mathrm{fund}})
\right),
\end{align}
where $\chi_{U(n_I)}^{\mathrm{adj}}$ is an adjoint character for the $U(n_I)$
and $\chi_{U(n_I)}^{\mathrm{fund}}(\chi_{U(n_I)}^{\overline{\mathrm{fund}}})$
is a (anti-) fundamental charater for the $U(n_I)$.
$i^{v}_I$ and $i^{h}_{I,I+1}$ each denote the single-particle
indices for the vector multiplet and the hyper multiplet.
They are explicitly given by \cite{arXiv:1911.10794,arXiv:2108.12090}
\footnote{In \cite{arXiv:2108.12090}, for the $S^5$ case the single-particle index for the hypermultiplet
was directly derived by analyzing the modes on D3-branes.
It is given by setting $w_{I+\frac{1}{2}}=q u_{I+2}$ in 
the definition of $i^{h}_{I,I+1}$ ($I=1,2,3$) in (\ref{ihyper}).
For the $T^{1,1}$ case we can obtain $i^{h}_{I,I+1}$ by simply replacing 
$w_{I+\frac{1}{2}}$ with the one for $T^{1,1}$ as discussed in \cite{arXiv:1911.10794}.}

\begin{align}
  i^{v}_I
  &=1-\frac{(1-q^{-3}w_{I-\frac{1}{2}}w_{I+\frac{1}{2}})(1-q^{\frac{3}{2}}y)(1-q^{\frac{3}{2}}y^{-1})}
  {(1-w_{I-\frac{1}{2}})(1-w_{I+\frac{1}{2}})},\\
  i^{h}_{I,I+1}
  &=\frac{w_{I+\frac{1}{2}}^{\frac{1}{2}}}{1-w_{I+\frac{1}{2}}}
  \frac{(1-q^{\frac{3}{2}}y)(1-q^{\frac{3}{2}}y^{-1})}{q^{\frac{3}{2}}}.
  \label{ihyper}
\end{align}
Also, for later convenience we denote the integrand in (\ref{d3integral})
as $F_{n_1,n_2,n_3,n_4}$:
\begin{align}
\label{Fconi}
F_{n_1,n_2,n_3,n_4}(q,y,u,v)=\int d\mu_1 d\mu_2 d\mu_3 d\mu_4 \Pexp[i_{\mathrm{all}}].
\end{align}

\subsection{Schur-like limit}
\label{Schursec}
As discussed in \cite{arXiv:2108.12090},
the determination of the contour in the gauge integral (\ref{Fconi})
requires special care.
To avoid the difficulty,
in this paper we instead 
consider an unrefined limit of the superconformal index
which we call the Schur-like limit.
\footnote{The Klebanov-Witten theory is related to $\mathrm{S}^5/\ZZ_2$ model
by mass deformation and this limit is related to the Schur limit of 
the index of that model.
Therefore we call this limit the ``Schur-like'' limit even though this theory
has only $\N=1$ symmetry.
}
Namely, we take the following specialization in the definition of the index
(\ref{defconiindex}):
\begin{align}
  v \rightarrow u,\qquad y\rightarrow 1,
  \end{align}
In this limit the single-particle indices 
$i^{h}_{2,3}$ and $i^{h}_{4,1}$ become simple form
\begin{align}
i^{h}_{2,3}=i^{h}_{4,1}=q^{-\frac{3}{4}}-q^{\frac{3}{4}}.
\end{align}
Then $F_{n_1,n_2,n_3,n_4}$ is factorized to
\begin{align}
F_{n_1,n_2,n_3,n_4}(q,u)=F_{n_1,n_2}(q,u) F_{n_3,n_4}(q,u^{-1})
q^{\frac{3}{2}(n_2 n_3+n_4 n_1)},
\end{align}
where $F_{n_1,n_2}(q,u)$ denotes the contribution of D3-branes with
wrapping numbers $(n_1,n_2,0,0)$ 
and given by $F_{n_1,n_2,0,0}(q,1,u,u)$ in (\ref{Fconi}).
Also, $F_{n_3,n_4}(q,u^{-1})$ corresponds to 
the contribution of $(0,0,n_3,n_4)$ configuration.
Moreover $F_{n_1,n_2}(q,u)$ is invariant under the exchange of the subscripts
\begin{align}
F_{n_1,n_2}(q,u)=F_{n_2,n_1}(q,u).
\end{align}

We calculated $F_{n_1,n_2}(q,u)$ up to $n_1+n_2 \leq 4$ and
here we show explicit results up to a certain order of $q$ which is
enough to reproduce all results in Section \ref{result.sec}.
\begin{align}
\label{coniF}
 F_{0,0}(q,u)&=1\nn
 F_{1,0}(q,u)&=\frac{u^2}{u^2-1}
 + (1-\frac{1}{u^2})q^{\frac{3}{2}}
 + (1-\frac{1}{u^6})q^3
 + (-\frac{1}{u^4}+\frac{1}{u^6}-\frac{1}{u^{10}}+1)q^{\frac{9}{2}}\nn
 &+ (1-\frac{1}{u^{14}})q^6
 +(-\frac{1}{u^6}+\frac{1}{u^{12}}-\frac{1}{u^{18}}+1)q^{\frac{15}{2}}
 + (\frac{1}{u^{10}}-\frac{1}{u^{12}}-\frac{1}{u^{22}}+1)q^9\nn
 &+(-\frac{1}{u^8}+\frac{1}{u^{18}}-\frac{1}{u^{26}}+1)q^{\frac{21}{2}}
 +\cdots,\nn
 F_{2,0}(q,u)&=\frac{u^6}{(u^2-1)^2 (u^2+1)}
 +q^{\frac{3}{2}}+ (\frac{1}{u^6}+2)q^3
 + (\frac{1}{u^4}+\frac{1}{u^{12}}+2)q^{\frac{9}{2}}\nn
 &+(\frac{3}{u^8}+\frac{1}{u^{18}}+3)q^6
 + (\frac{2}{u^{12}}+\frac{1}{u^{24}}+3)q^{\frac{15}{2}}
 + (\frac{1}{u^6}+\frac{2}{u^{10}}+\frac{4}{u^{16}}+\frac{1}{u^{30}}+4)q^9\nn
 &+(\frac{3}{u^{20}}+\frac{1}{u^{36}}+4)q^{\frac{21}{2}}
 +\cdots,\nn
 F_{1,1}(q,u)&=-\frac{ u^6 (u^2-3)q^{\frac{3}{2}}}{(u^2-1)^2 (u^2+1)}
 + (3-u^4)q^3
 + (-u^6+\frac{3}{u^6}+4)q^{\frac{9}{2}}\nn
 &+ (-u^8+\frac{1}{u^4}+\frac{3}{u^{12}}+5)q^6
 + (-u^{10}+\frac{6}{u^8}+\frac{3}{u^{18}}+6)q^{\frac{15}{2}}\nn
 &+(-u^{12}+\frac{3}{u^{12}}+\frac{3}{u^{24}}+7)q^9 
 +(-u^{14}+\frac{1}{u^6}+\frac{5}{u^{10}}+\frac{8}{u^{16}}+\frac{3}{u^{30}}+8)q^{\frac{21}{2}}
 +\cdots,\nn
 F_{3,0}(q,u)&=\frac{u^{12}}{(u^2-1)^3 (u^6+2 u^4+2 u^2+1)}
 +\frac{ u^4q^{\frac{3}{2}}}{u^4-1}
 +\frac{ (2 u^8+u^6-u^4+1)q^3}{u^4 (u^4-1)}\nn
 &+\frac{(3 u^{14}-3u^{12}+u^{10}+u^8-u^2+1)q^{\frac{9}{2}} }{u^{12} (u^2-1)}\nn
 &+\frac{ (4 u^{24}+u^{22}-2 u^{20}+u^{18}+2 u^{12}+u^{10}-u^4+1)q^6}
 {u^{20} (u^4-1)}\nn
 &+\frac{ (5 u^{32}-4u^{28}+2 u^{26}+u^{24}+u^{20}-2 u^{18}
 +2 u^{14}+u^{12}-u^4+1)q^{\frac{15}{2}}}{u^{28} (u^4-1)}
 +\cdots,\nn
 F_{2,1}(q,u)&=\frac{ u^{12} (u^6-2 u^4-2 u^2+6)q^3}
 {(u^2-1)^3 (u^2+1) (u^4+u^2+1)}
 +\frac{u^4 (u^{10}-2 u^6-2 u^4+6)q^{\frac{9}{2}} }{u^4-1}\nn
 &+\frac{(u^{22}-u^{18}-u^{16}-2 u^{14}+8 u^8+4 u^6-6 u^4+6)q^6}{u^4 (u^4-1)}\nn
 &-\frac{ (-u^{32}+u^{30}+u^{24}+u^{22}-u^{20}
 +u^{18}-14 u^{14}+14 u^{12}-4 u^{10}-4 u^8+6u^2-6)q^{\frac{15}{2}}}{u^{12} (u^2-1)}
 +\cdots,\nn
 F_{4,0}(q,u)&=\frac{u^{20}}{(u^2-1)^4 (u^2+1)^2
  (u^4+1) (u^4+u^2+1)}
  +\cdots,\nn
 F_{3,1}(q,u)&=-\frac{ u^{20} 
 (u^{12}-2 u^{10}-u^8+u^6+4 u^4+3 u^2-10)q^{\frac{9}{2}}}
 {(u^2-1)^4 (u^2+1)^2 (u^4+1) (u^4+u^2+1)}
 +\cdots,\nn
 F_{2,2}(q,u)&=\frac{ u^{20} (-3 u^{12}+5 u^{10}+4 u^8-3 u^6-9 u^4-8 u^2+20)q^6}
 {(u^2-1)^4 (u^2+1)^2 (u^4+1) (u^4+u^2+1)}
 +\cdots.
\end{align}

\section{Results and comparisons for $N=2$}\label{result.sec}
In this section we calculate the D3-branes contribution to
the indices and compare them with the gauge theory results
to confirm the formula (\ref{kwformula}).
In addition to the fugacity $q$ we expand the index in terms of $\zeta$,
\begin{align}
\mathcal{I^{\mathrm{KW}}}=\sum_{i\in \ZZ}\mathcal{I}^{\mathrm{KW}}_{B=i}\ \zeta^i,
\end{align}
and analyze each sector with each baryonic charge.

We show results for the $N=2$ with $B=0,1,2,3,4$ case. 
For the $B=0,2,4$ sectors we calculated the indices up to the leading contribution
of quadruple-wrapping D3-branes.
For the $B=1,3$ sectors which don't have the quadruple-wrapping D3-branes
contribution we calculated the indices up to $q^{12}$ terms. 
The gauge theory results are easily obtained by the localization method 
and shown in appendix \ref{gaugen2.sec}.
\subsection{$B=0$}
We first analyze the $B=0$ sector.
The corresponding D3-brane configurations for this sector are shown in Table \ref{b0config}.
We can easily calculate the D3-branes contribution to the index
by using the explicit results in (\ref{coniF}).
\begin{table}[htb]
  \caption{The D3-brane configurations for the $B=0$ sector are shown.
  A tuple $(n_1,n_2,n_3,n_4)$ denotes a configuration of $n_I$ D3-branes wrapped on each
  three-cycle $S_I$.}
  \vspace{-0.5cm}
  \begin{align*}
  \begin{array}{c|l|l|l|l}\hline
    B 	&\text{single}&\text{double} &\text{triple}&\text{quadruple}\\\hline
   \multirow{3}{*}{0}	&  & (n_1,n_2,n_3,n_4)=(1,1,0,0), &  &(2,2,0,0),(2,0,0,2),(0,2,2,0),  \\
   &  &(0,1,1,0),(0,0,1,1),(1,0,0,1) & &(0,0,2,2),(2,1,0,1),(1,2,1,0), \\	
   &  & & &(1,0,1,2),(0,1,2,1),(1,1,1,1) \\\hline
  \end{array}
  \end{align*}
  \label{b0config}
\end{table}

By summing up the contributions of all double-wrapping configurations in
Table \ref{b0config}, we get
\begin{align}
\mathcal{I}_{B=0}^{\mathrm{double}}&=\mathcal{I}_{(1,1,0,0)}+\mathcal{I}_{(0,1,1,0)}
+\mathcal{I}_{(0,0,1,1)}+\mathcal{I}_{(1,0,0,1)}\nn
&= \left(-\chi _2+3 \chi _4-\chi _6+3\right)q^{\frac{9}{2}}
+ \left(-5 \chi _2+3 \chi _4+\chi _6-\chi _8+2\right)q^6\nn
&+ \left(-3 \chi _2+4 \chi _4-2 \chi _6+\chi _8-\chi_{10}+5\right)q^{\frac{15}{2}}\nn
&+ \left(-9 \chi _2+7 \chi _4-3 \chi _6+3 \chi _8-\chi _{10}-\chi _{12}+10\right)q^9\nn
&-2 \left(4 \chi _2-4 \chi _4+\chi _6-2 \chi _8+\chi _{10}+\chi_{12}-5\right) q^{\frac{21}{2}}\nn
&+ \left(-7 \chi _2+7 \chi _4-5 \chi _6+5 \chi _8-4 \chi _{10}
+6 \chi _{12}-\chi _{14}-\chi _{16}-5 \chi _{18}+3 \chi _{20}+6\right)q^{12}
+\mathcal{O}(q^{\frac{25}{2}}),
\end{align}
where $\chi_n$ is $SU(2)$ character defined in
\begin{align}
\chi_n=\frac{u^{n+1}-u^{-(n+1)}}{u-u^{-1}}.
\end{align}
Also, the quadruple-wrapping D3-branes contribution is given by
\begin{align}
  \mathcal{I}_{B=0}^{\mathrm{quadruple}}&=\mathcal{I}_{(2,2,0,0)}+\mathcal{I}_{(2,0,0,2)}
  +\mathcal{I}_{(0,2,2,0)}+\mathcal{I}_{(0,0,2,2)}+\mathcal{I}_{(2,1,0,1)}
  +\mathcal{I}_{(1,2,1,0)}+\mathcal{I}_{(1,0,1,2)}+\mathcal{I}_{(0,1,2,1)}+\mathcal{I}_{(1,1,1,1)}\nn
  &=\left(-\chi _{2}+5 \chi _{4}-8 \chi _{6}+10 \chi _{8}-6 
  \chi _{12}-\chi _{14}+\chi _{16}+5 \chi _{18}-3 \chi _{20}+15\right)q^{12}
  +\mathcal{O}(q^{\frac{25}{2}}) ,
\end{align}
Then we can find the contribution of D3-branes in the $B=0$ sector
\begin{align}
  \mathcal{I}_{B=0}^{\mathrm{D}3}=1+\mathcal{I}_{B=0}^{\mathrm{double}}
  +\mathcal{I}_{B=0}^{\mathrm{quadruple}},
\end{align}
perfectly agrees with the gauge theory result (\ref{gaugeb0}). 
\footnote{The term ``1'' corresponds to $\mathcal{I}_{(0,0,0,0)}$.}
\subsection{$B=1$}
Next, we consider the $B=1$ sector.
The D3-brane configurations in this sector are summarized in Table \ref{b1config}.
\begin{table}[htb]
  \caption{The D3-brane configurations for the $B=1$ sector are shown.}
  \vspace{-0.5cm}
  \begin{align*}
  \begin{array}{c|l|l|l|l}\hline
    B 	&\text{single}&\text{double} &\text{triple}&\text{quadruple}\\\hline
   \multirow{2}{*}{1}	& (1,0,0,0),(0,0,1,0) &  & (2,1,0,0),(2,0,0,1),(0,1,2,0), &  \\
   &  & & (0,0,2,1), (1,1,1,0), (1,0,1,1) & \\\hline
  \end{array}
  \end{align*}
  \label{b1config}
\end{table}

We can easily calculate the single-wrapping D3 contribution,
\begin{align}
  \mathcal{I}_{B=1}^{\mathrm{single}}&=\mathcal{I}_{(1,0,0,0)}+\mathcal{I}_{(0,0,1,0)}\nn
  &= \chi _{2}q^{\frac{3}{2}}+ \left(\chi _{2}-3\right)q^3
  +\left(2 \chi _{2}-\chi _{4}-1\right)q^{\frac{9}{2}} 
  +\left(-\chi _{2}+\chi _{4}+\chi _{6}-\chi _{8}\right)q^6 \nn
  &+\left(\chi _{2}+\chi _{10}-\chi_{12}-1\right)q^{\frac{15}{2}} 
  +\left(2 \chi _{2}-\chi _{4}-\chi _{8}+\chi _{10}+\chi _{14}-\chi _{16}-1\right)q^9 \nn
  &+ \left(\chi _{2}-\chi _{6}+2 \chi _{8}-\chi _{10}
  +\chi _{18}-\chi_{20}-1\right)q^{\frac{21}{2}}\nn
 & + \left(\chi _{2}+\chi _{4}-\chi _{6}-\chi _{14}+\chi _{16}+\chi _{22}
  -\chi _{24}-1\right)q^{12}
  +\mathcal{O}(q^{\frac{25}{2}}) ,
  \end{align}
and also the triple-wrapping D3s contribution is given by
\begin{align}
  \mathcal{I}_{B=1}^{\mathrm{triple}}&=\mathcal{I}_{(2,1,0,0)}+\mathcal{I}_{(2,0,0,1)}
  +\mathcal{I}_{(0,1,2,0)}+\mathcal{I}_{(0,0,2,1)}
  +\mathcal{I}_{(1,1,1,0)}+\mathcal{I}_{(1,0,1,1)}\nn
  &=q^{\frac{15}{2}} \left(4 \chi _{2}-3 \chi _{4}+5 \chi _{6}-\chi _{8}-
  2 \chi _{10}+\chi _{12}+1\right)\nn
 & + \left(10 \chi _{2}-9 \chi _{4}+6 \chi _{6}+\chi _{8}
  -\chi _{10}-\chi _{12}-\chi _{14}+\chi_{16}-14\right)q^9\nn
  &+ \left(12 \chi _{2}-9 \chi _{4}+9 \chi _{6}-4 \chi _{8}+2 \chi _{10}
  -\chi _{12}-\chi _{14}-\chi _{18}+\chi _{20}-14\right)q^{\frac{21}{2}}\nn
  &+ \left(14 \chi _{2}-9 \chi _{4}+8\chi _{6}-4 \chi _{8}
  +3 \chi _{10}-2 \chi _{12}-\chi _{16}-\chi _{22}+\chi _{24}-9\right)q^{12}
  +\mathcal{O}(q^{\frac{25}{2}}).      
\end{align}
We can find the sum of these contributions
\begin{align}
  \mathcal{I}_{B=1}^{\mathrm{D}3}&
  =\mathcal{I}_{B=1}^{\mathrm{single}}+\mathcal{I}_{B=1}^{\mathrm{triple}}
  \end{align}
coincides with the gauge theory result (\ref{gaugeb1}).
\subsection{$B=2$}
Then we consider the $B=2$ sector.
The D3-brane configurations are shown in Table \ref{b2config}.
\begin{table}[htb]
  \caption{The D3-brane configurations for the $B=2$ sector are shown.}
  \vspace{-0.5cm}
  \begin{align*}
  \begin{array}{c|l|l|l|l}\hline
    B 	&\text{single}&\text{double} &\text{triple}&\text{quadruple}\\\hline
   \multirow{3}{*}{2}	&  & (2,0,0,0),(0,0,2,0),(1,0,1,0) &  & (3,1,0,0), (3,0,0,1), (0,1,3,0)\\
   &  & & & (0,0,3,1) ,(2,0,1,1),(2,1,1,0) \\	
   &  & & &(1,1,2,0), (1,0,2,1)\\\hline
  \end{array}
  \end{align*}
  \label{b2config}
\end{table}

We can get the double-wrapping contribution,
\begin{align}
  \mathcal{I}_{B=2}^{\mathrm{double}}&=\mathcal{I}_{(2,0,0,0)}+\mathcal{I}_{(0,0,2,0)}
  +\mathcal{I}_{(1,0,1,0)}\nn
  &=  \left(\chi _{4}+1\right)q^3
  + \left(-2 \chi _{2}+\chi _{4}+1\right)q^{\frac{9}{2}}
  + \left(-2 \chi _{2}+2 \chi _{4}-\chi _{6}+3\right)q^6\nn
  &+ \left(-4 \chi _{2}+3 \chi _{4}-\chi _{6}+\chi_{8}-\chi _{10}+6\right)q^{\frac{15}{2}}\nn
  &+ \left(-4 \chi _{2}+4 \chi _{4}-\chi _{6}+2 \chi _{8}-\chi _{10}-\chi _{12}+5\right)q^9\nn
  &+\left(-3 \chi _{2}+3 \chi _{4}-3 \chi _{6}+3 \chi _{8}
  -2 \chi_{10}+3 \chi _{12}-\chi _{14}-2 \chi _{18}+\chi _{20}+3\right)q^{\frac{21}{2}} 
  +\mathcal{O}(q^{\frac{23}{2}}) ,
  \end{align}
and also the quadruple-wrapping contribution
  \begin{align}
    \mathcal{I}_{B=2}^{\mathrm{quadruple}}&=\mathcal{I}_{(3,1,0,0)}+\mathcal{I}_{(3,0,0,1)}
    +\mathcal{I}_{(0,1,3,0)}+\mathcal{I}_{(0,0,3,1)}+\mathcal{I}_{(2,0,1,1)}
    +\mathcal{I}_{(2,1,1,0)}+\mathcal{I}_{(1,1,2,0)}+\mathcal{I}_{(1,0,2,1)}\nn
    &= \left(3 \chi _{4}-3 \chi _{6}+5 \chi _{8}-3 \chi _{12}+2 \chi _{18}-\chi _{20}+7\right)q^{\frac{21}{2}}
    +\mathcal{O}(q^{\frac{23}{2}}).
  \end{align}  
The sum of these contributions
\begin{align}
\mathcal{I}_{B=2}^{\mathrm{D}3}
&=\mathcal{I}_{B=2}^{\mathrm{double}}+\mathcal{I}_{B=2}^{\mathrm{quadruple}}
\end{align}
is consistent with the gauge theory result (\ref{gaugeb2}).
\subsection{$B=3$}
Next, let us consider the $B=3$ sector.
The brane configurations are shown in Table \ref{b3config}.
\begin{table}[htb]
  \caption{The D3-brane configurations for the $B=3$ sector are shown.}
  \vspace{-0.5cm}
  \begin{align*}
  \begin{array}{c|l|l|l|l}\hline
    B 	&\text{single}&\text{double} &\text{triple}&\text{quadruple}\\\hline
   3	&  &  & (3,0,0,0),(0,0,3,0),(2,0,1,0),(1,0,2,0) &  \\\hline
  \end{array}
  \end{align*}
  \label{b3config}
\end{table}

We obtain the follwing result:
\begin{align}
  \mathcal{I}_{B=3}^{\mathrm{D}3}&= \mathcal{I}_{B=3}^{\mathrm{triple}}=\mathcal{I}_{(3,0,0,0)}+\mathcal{I}_{(0,0,3,0)}
  +\mathcal{I}_{(2,0,1,0)}+\mathcal{I}_{(1,0,2,0)}\nn
  &= \left(\chi _{2}+\chi _{6}\right)q^{\frac{9}{2}}
  + \left(2 \chi _{2}-2 \chi _{4}+\chi _{6}-3\right)q^6
  + \left(3 \chi _{2}-2 \chi _{4}+2 \chi _{6}-\chi _{8}-4\right)q^{\frac{15}{2}}\nn
  &+ \left(5 \chi _{2}-3 \chi_{4}+2 \chi _{6}-\chi _{8}+\chi _{10}-\chi _{12}-3\right)q^9\nn
  &+ \left(7 \chi _{2}-5 \chi _{4}+8 \chi _{6}-4 \chi _{8}
  +2 \chi _{10}-\chi _{12}-\chi _{14}-10\right)q^{\frac{21}{2}}\nn
  &+ \left(18 \chi_{2}-13 \chi _{4}+7 \chi _{6}-4 \chi _{8}
  +5 \chi _{10}-2 \chi _{12}-\chi _{16}-14\right)q^{12}
  +\mathcal{O}(q^{\frac{25}{2}}).
\end{align}
This result agrees with the gauge theory result (\ref{gaugeb3}).

\subsection{$B=4$}
Finally, we consider the $B=4$ sector.
We show the D3-brane configurations for this sector in Table \ref{b4config}.
\begin{table}[htb]
  \caption{The D3-brane configurations for the $B=4$ sector are shown.}
  \vspace{-0.5cm}
  \begin{align*}
  \begin{array}{c|l|l|l|l}\hline
    B 	&\text{single}&\text{double} &\text{triple}&\text{quadruple}\\\hline
   \multirow{2}{*}{4}	&  &  &  & (4,0,0,0), (0,0,4,0), (3,0,1,0),\\
   &  & & & (1,0,3,0), (2,0,2,0)\\\hline
  \end{array}
  \end{align*}
  \label{b4config}
\end{table}

We get the following quadruple-wrapping D3-branes contribution
\begin{align}
  \mathcal{I}_{B=4}^{\mathrm{D}3}&=\mathcal{I}_{B=4}^{\mathrm{quadruple}}=\mathcal{I}_{(4,0,0,0)}+\mathcal{I}_{(0,0,4,0)}
  +\mathcal{I}_{(3,0,1,0)}+\mathcal{I}_{(1,0,3,0)}+\mathcal{I}_{(2,0,2,0)}\nn
  &= \left(\chi _{4}+\chi _{8}+1\right)q^6
  +\mathcal{O}(q^{\frac{13}{2}}),
\end{align}  
and this result agrees with (\ref{gaugeb4}).

\section{Conclusions and discussion}\label{disc.sec}
In this paper we calculated the Schur-like index of the Klebanov-Witten theory
from our proposing formula (\ref{kwformula}) based on the AdS/CFT correspondence.
We numerically checked the matching between the index of the gravity side
and the gauge theory side for $N=2$ regarding each baryonic charge sector.
Especially for $B=0,2,4$ we calculated the index up to the first order term of
the contribution of quadruple-wrapping D3-branes.
For $B=1,3$ we calculated up to $q^{12}$ order terms.
In every case we found perfect agreement with the gauge theory results.

There are many future directions.
Our analysis is limited to the Schur-like index.
Hence it is the remaining task to check if the giant graviton expansion works for the general superconformal index.
Also, Our formula (\ref{kwformula}) has multiple sums of the wrapped D3-brane contributions,
and it is an interesting problem to discuss whether we can reduce the multiple sums to the simple sum 
discussed in \cite{arXiv:2109.02545} by suitably choosing the expansion variable.
We will return to this point in the near future.
It is also an interesting problem to check the
giant graviton expansion (\ref{toricformula}) including multiple-wrapping D3-branes for other 
Sasaki-Einstein manifolds.
Furthermore, such giant graviton expansion could be applied to 
a much broader range of the AdS/CFT correspondence. 
In particular, its application to theories with no Lagrangian description is a very important issue.

\section*{Acknowledgments}
The author would like to give his great thanks to Yosuke Imamura for
collaboration in the early stage of this work and excellent advice.
The author also thanks Daisuke Yokoyama, Shuichi Murayama, and Tatsuya Mori for helpful discussions.
SF is supported by the South African Research Chairs Initiative of the Department of Science and Innovation
and the National Research Foundation grant 78554.

\appendix
\section{Results from gauge theory calculation}
\label{gaugen2.sec}
We show the results of the ratio of the indices of the gauge theory side and the
the Kalzua-Klein index for $N=2$.
\begin{align}
\label{gaugeb0}
\frac{\mathcal{I}^{\mathrm{KW}}_{B=0}}{\mathcal{I}_{\mathrm{KK}}^{T^{1,1}}}
&= \left(-\chi _2+3 \chi _4-\chi _6+3\right)q^{\frac{9}{2}}
+ \left(-5 \chi _2+3 \chi _4+\chi _6-\chi _8+2\right)q^6\nn
&+ \left(-3 \chi _2+4 \chi _4-2 \chi _6+\chi _8-\chi_{10}+5\right)q^{\frac{15}{2}}\nn
&+ \left(-9 \chi _2+7 \chi _4-3 \chi _6+3 \chi _8-\chi _{10}-\chi _{12}+10\right)q^9\nn
&-2 \left(4 \chi _2-4 \chi _4+\chi _6-2 \chi _8+\chi _{10}+\chi_{12}-5\right) q^{\frac{21}{2}}\nn
&+ \left(-8 \chi_2+12 \chi_4-13 
\chi_6+15 \chi_8-4 \chi_{10}-2 \chi_{14}+21\right)q^{12}
+\mathcal{O}(q^{\frac{25}{2}}),
\end{align}
\begin{align}
  \label{gaugeb1}
  \frac{\mathcal{I}^{\mathrm{KW}}_{B=1}}{\mathcal{I}_{\mathrm{KK}}^{T^{1,1}}}
&= \chi _{2}q^{\frac{3}{2}}
 +\left(\chi _{2}-3\right)q^3
+ \left(2 \chi _{2}-\chi _{4}-1\right)q^{\frac{9}{2}}
+ \left(-\chi _{2}+\chi _{4}+\chi _6-\chi _{8}\right)q^6\nn
&+ \left(5 \chi _{2}-3 \chi _{4}+5\chi _6-\chi _{8}-\chi _{10}\right)q^{\frac{15}{2}}\nn
&+ \left(12 \chi _{2}-10 \chi _{4}+6 \chi _6-\chi _{12}-15\right)q^9\nn
&+ \left(13 \chi _{2}-9 \chi _{4}+8 \chi _6-2 \chi _{8}+\chi _{10}-\chi
_{12}-\chi _{14}-15\right)q^{\frac{21}{2}}\nn
&+\left(15 \chi _{2}-8 \chi _{4}+7 \chi _6-4 \chi _{8}+3 \chi _{10}
-2 \chi _{12}-\chi _{14}-10\right)q^{12} 
+\mathcal{O}(q^{\frac{25}{2}}),
\end{align}
\begin{align}
    \label{gaugeb2}
    \frac{\mathcal{I}^{\mathrm{KW}}_{B=2}}{\mathcal{I}_{\mathrm{KK}}^{T^{1,1}}}
    &=\left(\chi _{4}+1\right)q^3
  + \left(-2 \chi _{2}+\chi _{4}+1\right)q^{\frac{9}{2}}
  + \left(-2 \chi _{2}+2 \chi _{4}-\chi _{6}+3\right)q^6\nn
  &+ \left(-4 \chi _{2}+3 \chi _{4}-\chi _{6}+\chi_{8}-\chi _{10}+6\right)q^{\frac{15}{2}}\nn
  &+ \left(-4 \chi _{2}+4 \chi _{4}-\chi _{6}+2 \chi _{8}-\chi _{10}-\chi _{12}+5\right)q^9\nn
  &+ \left(-3 \chi _{2}+6 \chi _{4}-6 \chi _6+8 \chi _{8}-2 \chi
  _{10}-\chi _{14}+10\right)q^{\frac{21}{2}}
  +\mathcal{O}(q^{\frac{23}{2}}) ,
\end{align}
\begin{align}
  \label{gaugeb3}
  \frac{\mathcal{I}^{\mathrm{KW}}_{B=3}}{\mathcal{I}_{\mathrm{KK}}^{T^{1,1}}}
  &= \left(\chi _{2}+\chi _{6}\right)q^{\frac{9}{2}}
  + \left(2 \chi _{2}-2 \chi _{4}+\chi _{6}-3\right)q^6
  + \left(3 \chi _{2}-2 \chi _{4}+2 \chi _{6}-\chi _{8}-4\right)q^{\frac{15}{2}}\nn
  &+ \left(5 \chi _{2}-3 \chi_{4}+2 \chi _{6}-\chi _{8}+\chi _{10}-\chi _{12}-3\right)q^9\nn
  &+ \left(7 \chi _{2}-5 \chi _{4}+8 \chi _{6}-4 \chi _{8}
  +2 \chi _{10}-\chi _{12}-\chi _{14}-10\right)q^{\frac{21}{2}}\nn
  &+ \left(18 \chi_{2}-13 \chi _{4}+7 \chi _{6}-4 \chi _{8}
  +5 \chi _{10}-2 \chi _{12}-\chi _{16}-14\right)q^{12}
  +\mathcal{O}(q^{\frac{25}{2}}) ,  
\end{align}
\begin{align}
  \label{gaugeb4}
  \frac{\mathcal{I}^{\mathrm{KW}}_{B=4}}{\mathcal{I}_{\mathrm{KK}}^{T^{1,1}}}
  &= \left(\chi _{4}+\chi _{8}+1\right)q^6
  +\mathcal{O}(q^{\frac{13}{2}}),
\end{align}

\end{document}